\documentclass[%
reprint,
superscriptaddress,
amsmath,amssymb,
aps,
floatfix,
]{revtex4-1}

\usepackage{graphicx}% Include figure files
\usepackage{subcaption}
\usepackage{dcolumn}% Align table columns on decimal point
\usepackage{bm}% bold math
\usepackage{braket}
\usepackage{amsmath}
\usepackage{mathtools}
\usepackage[colorlinks,allcolors=blue]{hyperref} 
\usepackage{hyperref}% add hypertext capabilities
\newcommand{\expect}[1]{\left\langle #1 \right\rangle}

\begin{document}
	
	\raggedbottom
	
	\preprint{APS/123-QED}
	
	\title{Microwave Photon Number Resolving Detector Using the Topological Surface State of Superconducting Cadmium Arsenide}
	
	\author{Eric Chatterjee}
	\affiliation{Sandia National Laboratories, Livermore, California 94550, USA}
	\author{Wei Pan}
	\affiliation{Sandia National Laboratories, Livermore, California 94550, USA}
	\author{Daniel Soh}
	\affiliation{Sandia National Laboratories, Livermore, California 94550, USA}

	\date{\today}% It is always \today, today,
	
	\begin{abstract}
		Photon number resolving detectors play a central role in quantum optics. A key challenge in resolving the number of absorbed photons in the microwave frequency range is finding a suitable material that provides not only an appropriate band structure for absorbing low-energy photons but also a means of detecting a discrete photoelectron excitation. To this end, we propose to measure the temperature gain after absorbing a photon using superconducting cadmium arsenide (Cd\textsubscript{3}As\textsubscript{2}) with a topological semimetallic surface state as the detector.  The surface electrons absorb the incoming photons and then transfer the excess energy via heat to the superconducting bulk's phonon modes. The temperature gain can be determined by measuring the change in the zero-bias bulk resistivity, which does not significantly affect the lattice dynamics. Moreover, the obtained temperature gain scales discretely with the number of absorbed photons, enabling a photon-number resolving function. Here, we will calculate the temperature increase as a function of the number and frequency of photons absorbed. We will also derive the timescale for the heat transfer process from the surface electrons to the bulk phonons. We will specifically show that the transfer processes are fast enough to ignore heat dissipation loss. 
	\end{abstract}
	
	\pacs{Valid PACS appear here}% PACS, the Physics and Astronomy
	\maketitle
	
\section{Introduction} \label{sec: Introduction}

Photon number resolving detectors have been explored significantly over the past decades \cite{divochiy2008superconducting, kardynal2008avalanche, rosenberg2005noise} due to the dire need for resolving the number of photons in applications such as the security of quantum communications \cite{deng2005improving, chen2013multi} and the sensitivity of quantum sensing \cite{afek2010high}. As photon-based quantum computing advances, precise resolution of photon number detection is increasingly important. Microwave photons are the backbone of prolific transmon quantum computation, and therefore, detection of the microwave photons is tremendously important in the current quantum computing paradigm \cite{houck2008controlling}. Several non-number-resolving techniques to detect microwave photons have been developed, including the circuit QED technique \cite{romero2009microwave}, dressed-state superconducting quantum circuit \cite{inomata2016single}, current-biased Josephson junction \cite{poudel2012quantum}, and the dark-state detector \cite{royer2018itinerant}. It is well-known that building a parallel detection system, first splitting the light path using beam splitters and then using non-number-resolving detectors in each parallel path, may provide a probabilistic photon number resolving detection, which is further limited due to the loss associated with parallelization. In contrast, a single photon-number resolving detector with a deterministic photon number resolution would provide a immense advantage particularly in photonic quantum computers by reducing the error-correcting overhead. To the best of our knowledge, a single-device photon-number resolving detector that can simultaneously detect multiple incoming photons at microwave frequency has not been reported so far.  

Here, we propose a photon-number resolving detector operating at microwave frequencies, based on the topological surface states of cadmium arsenide (Cd$_3$As$_2$). Semimetals such as graphene provide an ideal detecting material for microwave photons due to their zero band gap. Recently, Dirac and Weyl semimetals with Dirac cone dispersion have gained prominence due to high mobility \cite{liang2015mobility}, along with the fact that they can be synthesized through conventional techniques \cite{AliCd3As2Structure,schumannQHE2018,uchidaQH2017}. Particularly, Cd$_3$As$_2$ displays proximity-induced bulk superconductivity at low temperatures, and the electronic structures of the bulk and the topological surface states are decoupled. Maintaining the Cd$_3$As$_2$ semimetal material at a very low temperature is necessary for an efficient photon-induced electron excitation to a conduction band just above the Fermi level due to the low photon energy. The bulk state enters a superconducting state at a sufficiently low temperature, opening a band gap beyond the microwave photon energy. Fortunately, the topological surface state of Cd$_3$As$_2$ is not affected by the temperature, continuing to provide a gapless Dirac cone. We use this topological surface state as a photon absorber. Once the photon is absorbed, a rapid rethermalization in band population occurs with a new elevated temperature corresponding to the absorbed photon energy. We then utilize the fact that the redistributed electron population transfers its energy to the bulk's phonon modes via a surface electron-bulk phonon coupling, thus increasing the bulk's temperature. The elevated bulk temperature then reduces the conductance of the superconducting bulk electron state, which is measured and used to eventually indicate the number of photons absorbed.

The paper is organized as follows. In Sec.~\ref{sec: Overview}, we briefly review the photon absorption in the topological surface state of Cd$_3$As$_2$.  In Sec.~\ref{sec: Temperature Increase vs. Absorbed Photon Number}, we connect the event of photon absorption in the topological surface state electrons to the bulk temperature increase via a two-step process, namely, the energy gain for surface electron modes and, then, the transfer of electron energy to the bulk phonon modes. Section~\ref{sec: Electron-Phonon Interaction Timescale} resolves the important issue of the time scale of the temperature increase and shows that the absorbed photon energy indeed is most likely transferred to the increase of the bulk temperature, rather than being lost to radiative decay of the excited electrons. Section~\ref{sec: Discussion} presents the construction of a photon-number resolving detector based on the results of the previous sections. Numerical examples with realistic parameters build credible real use cases of the proposed scheme. In the final section, we summarize our results and suggest the path toward building a photon-number resolving detector with near-unity efficiency on a chip.

\section{Photon absorption in topological surface state} \label{sec: Overview}

\begin{figure}[!tb]
	\centering
	\includegraphics[width=\linewidth]{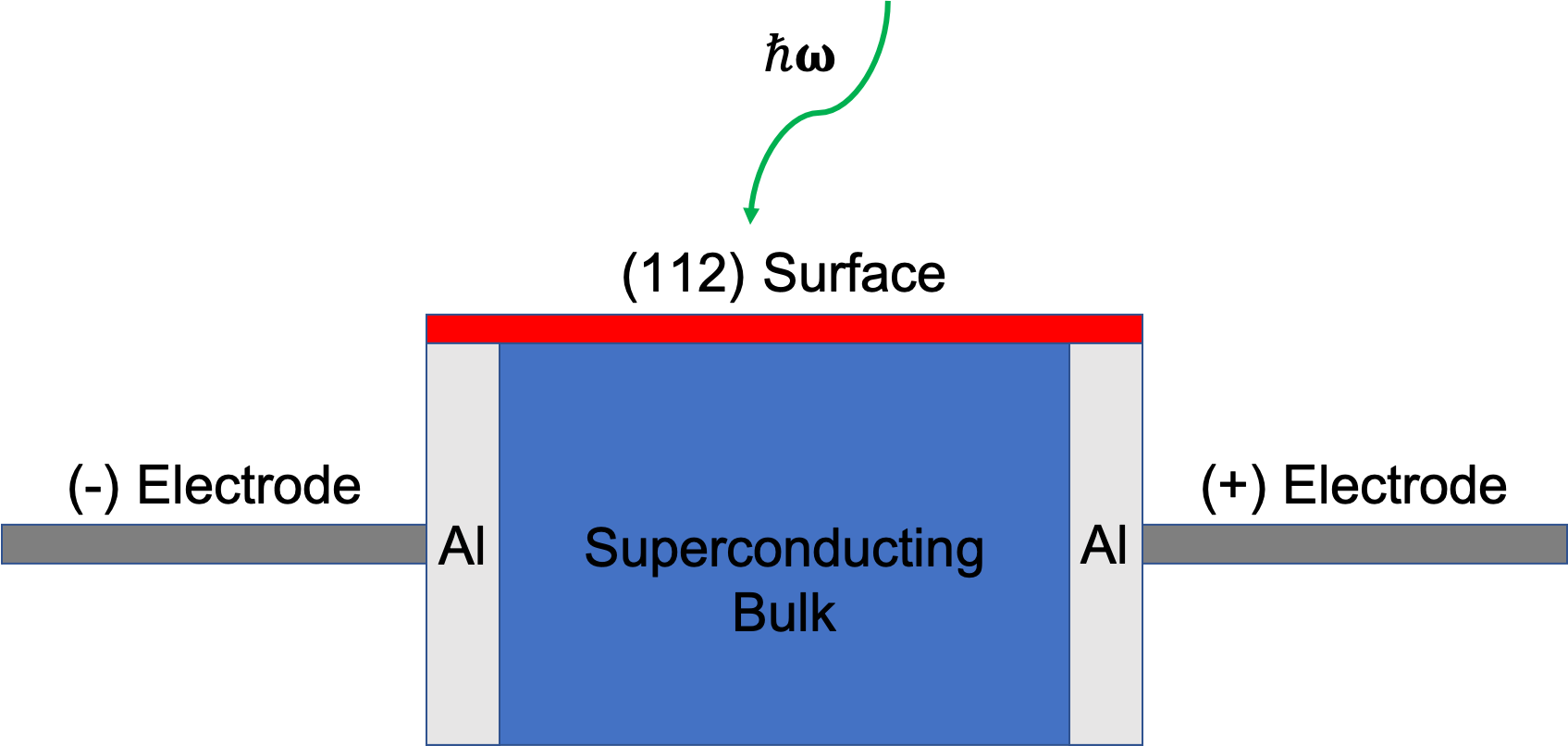}
	\caption{Basic layout for the Cd\textsubscript{3}As\textsubscript{2} photon number resolving device. At low temperatures, the bulk becomes superconducting due to the material's proximity to superconducting aluminum, while the (112) surface retains a graphene-like dispersion. A photon (depicted by the arrow) is absorbed by the surface electrons. The change in bulk resistivity (measured by the electrodes at zero bias) is used to determine the temperature increase. Note that the Al bars are positioned in the out-of-plane direction with respect to the Cd\textsubscript{3}As\textsubscript{2} material.}
	\label{fig:devicediagram}
\end{figure}

\begin{figure}[!tb]
	\centering
	\includegraphics[width=0.7\linewidth]{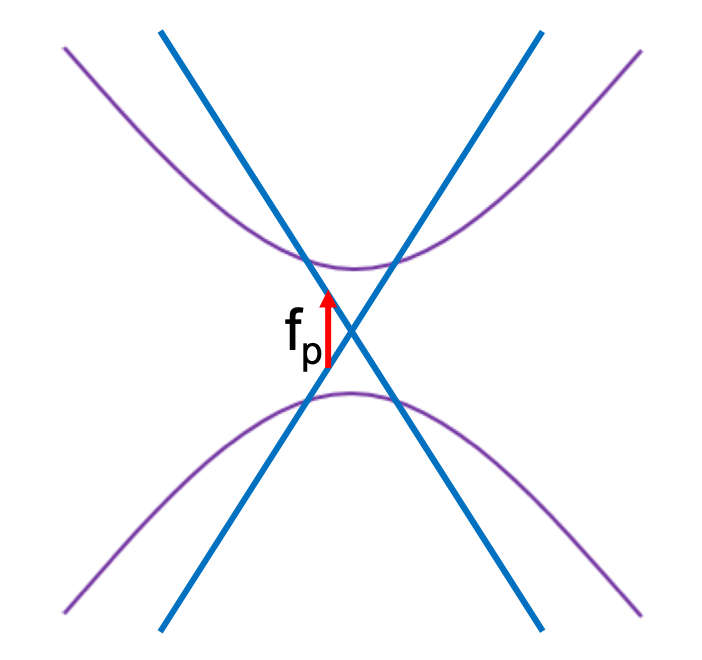}
	\caption{Band structures for the superconducting bulk (purple curves) and the topological surface state (blue straight lines). Note that a gap of frequency $f_\mathrm{gap}$ is opened in the bulk, while the surface state remains gapless, thus restricting the absorption of photons at microwave frequency $f_p$ to the surface state.}
	\label{fig:bandscomparison}
\end{figure}

The setup for the device is depicted in Fig.~\ref{fig:devicediagram}.
The system is based on a Cd\textsubscript{3}As\textsubscript{2} crystal inside a low-temperature refrigerator with a baseline temperature below the bulk superconducting critical temperature. Recent experimental findings have demonstrated that Cd\textsubscript{3}As\textsubscript{2} features a topological surface state on the (112) surface with a linear band crossing around the Dirac points \cite{YiTSS, GoyalTSS}. This graphene-like surface state band structure at low energy can be attributed to the fact that sites consisting of stacked As and Cd atoms approximately form a honeycomb superlattice on the (112) surface \cite{AliCd3As2Structure}. As with graphene, the dispersion relationship can be expressed as the following linear function of the wavevector $\bm{k}$ when the Fermi level is at the Dirac point:
\begin{equation} \label{eq: electron dispersion}
E_{c,v}(\bm{k}) = \pm \hbar v_F |\bm{k}|,
\end{equation}
where $\hbar$ is the Planck constant, $c$ and $v$ represent the conduction and valence bands, respectively, and $v_F$ denotes the Fermi velocity, which is approximately $10^6 \textrm{ m/s}$ \cite{YiTSS, NeupaneSemimetal}. At low temperatures, if the material is in proximity to superconducting aluminum (Al), the surface state electrons are decoupled from the bulk state, and the bulk becomes superconducting below 0.7 K while the surface retains its semimetallic property \cite{SuperconductingBulkVsTSS}. We measure the superconducting gap frequency $f_\mathrm{gap}$ near zero temperature ($T = 21 \textrm{ mK}$) and at $T = 0.39 \textrm{ K}$ from Figs. 2(a) and 1(d) of Yu \textit{et al.} \cite{SuperconductingBulkVsTSS}. For the case of near-zero temperature, the gap is measured as 0.113 meV (corresponding to a frequency of 27 GHz), while at 0.39 K, the gap amounts to 0.088 meV (corresponding to 21 GHz). The resulting band structures for the superconducting bulk and the topological surface state are compared in Fig.~\ref{fig:bandscomparison}. 
Note that this proximity-induced superconducting gap is smaller than the gap predicted from BCS theory \cite{BCSSuperconductivity}, which would equal about 50 GHz for a critical temperature around 0.7 K. Nonetheless, even for temperatures as high as 0.35-0.45 K, photons of microwave frequency below approximately 20 GHz will be absorbed solely by the surface state. Since we are primarily interested in microwave photons of frequency 5-10 GHz, this satisfies our goal of using the surface as the absorber and the bulk as the thermometer.
\par
In order to derive the absorption probability, we consider the physical manifestation of the Dirac cone on the nature of the Bloch states. As in graphene, each electronic state in the vicinity of a Dirac point can be conceptualized as a massless Dirac fermion with a well-defined momentum $\hbar \bm{k}$, where $\bm{k}$ represents the wavevector of the electronic state in the reciprocal space for which the Dirac point is the origin. Therefore, the absorption coefficient for Cd\textsubscript{3}As\textsubscript{2} will equal the corresponding value for graphene \cite{GrapheneAbsorption}:
\begin{align} \label{eq: absorption rate}
\begin{split}
A(\omega) &= \frac{e^2}{4\hbar\epsilon_0 c} \Bigg(f\bigg(-\frac{\hbar\omega}{2},T\bigg) - f\bigg(\frac{\hbar\omega}{2},T\bigg)\Bigg) \\
&= \frac{e^2}{4\hbar\epsilon_0 c} \tanh{\bigg(\frac{\hbar\omega}{4k_B T}\bigg)},
\end{split}
\end{align}
where $e$ is the charge of an electron, $\epsilon_0$ is the vacuum permittivity, $c$ the speed of light, and $f(E,T)$ denotes the Fermi-Dirac electron occupation probability at energy $E$ for temperature $T$. Note that the absorption coefficient (as a function of photon frequency) is invariant with respect to the Fermi velocity. This is because the interband dipole matrix element (corresponding to the absorption probability for a single electronic state) increases with the Fermi velocity, while the density of states decreases with it, eventually canceling each other's effect. At high frequencies ($\hbar\omega >> k_B T$), the absorption coefficient (i.e., the quantum efficiency for a single pass through a single Cd\textsubscript{3}As\textsubscript{2} crystal) is approximately invariant with frequency, equaling a constant value of $e^2/(4\hbar\epsilon_0 c) = 2.3 \%$. On the other hand, at low frequencies on the scale $\hbar \omega \lessapprox k_B T$, the coefficient becomes attenuated, reaching a minimum value of zero as the Dirac cone is approached. We will use an upper bound baseline temperature of 0.45 K, which will set the minimum quantum efficiency for photons of a given frequency interacting with a single crystal. For microwave frequencies ranging from 5-10 GHz, Eq.~\eqref{eq: absorption rate} implies a single-crystal quantum efficiency ranging from 0.3-0.6\%.
\par

\section{Temperature Increase vs. Absorbed Photon Number} \label{sec: Temperature Increase vs. Absorbed Photon Number}

Having determined the probability that the topological surface state absorbs a photon from an incoming field, our next step is to determine how the absorption of a single photon increases the temperature of the 3D bulk sample. When a photon excites an electron to the conduction band, a rapid rethermalization of the Fermi-Dirac distribution through electron-electron interaction ensues, leading to an electron temperature above the lattice temperature. For undoped monolayer graphene, which features a band structure approximately identical to that for the Cd$_3$As$_2$ surface state, this process occurs in tens of picoseconds for cryogenic baseline temperatures \cite{MihnevCarrierRelaxationTime}. The carrier rethermalization is followed by heat transfer from the collection of electrons to the lattice via electron-acoustic phonon interaction, until a thermal equilibrium is reached between the electron temperature and the lattice temperature. Generally, the interaction between electrons and acoustic phonons is much slower than the electron-electron interaction \cite{UltrafastDiracFermionRelaxation,GrapheneUltrafastCarrierDynamics,LundgrenFiete}. Afterward, the Fermi-Dirac distribution for the electron bands and the Bose-Einstein distribution for the phonon branches will comply with the same temperature.
\par 
We now determine the temperature increase due to the absorption of a photon of frequency $\omega$ by calculating the portion of the imparted energy that is eventually converted to bulk lattice vibrations (i.e., the phonon modes) and to the surface electron modes, and by deriving the heat capacity of the two systems. We will make two important assumptions here: first, that the electron-electron interaction rate dominates over the radiative loss rate of the electrons (which we will demonstrate in a later section), and second, that the very low values for the initial and final temperatures ensure that virtually all of the phonons are located in the low-energy linear parts of the acoustic branches, thus allowing for use of the Debye approximation \cite{Debye} in determining the heat capacity. 

\subsection{Energy Gain for Surface Electron Modes}
We start by writing out an expression corresponding to energy conservation in the system given the absorption of a photon of frequency $\omega$:
\begin{equation} \label{eq: energy conservation}
\hbar \omega = \Delta U_{el} + \Delta U_{ph},
\end{equation}
where $\Delta U_{el}$ and $\Delta U_{ph}$ represent the total energy gained by the topological surface electron modes and the bulk phonon modes, respectively, at equilibrium. We focus first on the energy gain for the electron modes as a function of electron temperature, as this will be necessary for calculating the initial electron temperature gain after photon absorption but prior to heat transfer to the lattice. The total electron energy with respect to the Fermi sea is calculated by taking a sum of the conduction and valence band energies weighted by the Fermi-Dirac occupation probabilities (multiplied by 2 to account for the fact that each spatial state contains 2 spin states):
\begin{equation}
U_{el}(T) = 2 \sum_{\bm{k}} \bigg(E_{c,\bm{k}} f(E_{c,\bm{k}},T) - E_{v,\bm{k}} \Big(1 - f(E_{v,\bm{k}},T)\Big)\bigg).
\end{equation}
The first term corresponds to the energy gained in creating a conduction band electron, while the second term corresponds to the energy gained in creating a valence band hole. Since the deviation from the Fermi sea at low temperatures is concentrated in the vicinity of the Dirac cone, we can assume that the linear isotropic dispersion relationship in Eq.~\eqref{eq: electron dispersion} holds for the entire relevant wavevector range. Therefore, the summation over wavevectors can be replaced by an integral over the density of spatial states ($\rho_c$ and $\rho_v$ for the conduction and valence bands, respectively):
\begin{align}
\begin{split}
U_{el}(T) &= 2 \int_{0}^{-\infty} dE_v \rho_v(E_v) (-E_v) \Bigg(1 - \frac{1}{e^{\frac{E_v}{k_B T}} + 1}\Bigg) \\
&\quad + 2 \int_{0}^{\infty} dE_c \rho_c(E_c) E_c \frac{1}{e^{\frac{E_c}{k_B T}} + 1}.
\end{split}
\end{align}
Since the conduction and valence band energies are opposite at each wavevector, we can express $U_{el}$ as a single integral over the energy absolute value $E$, where $E_c = E$ and $E_v = -E$. Due to the equivalent magnitudes of the dispersion slope for the conduction and valence bands at each wavevector, we can further define a general density of states $\rho(E) = \rho_v(E) = \rho_c(E)$:
\begin{align} \label{eq: electron energy increase integral}
\begin{split}
U_{el}(T) &= 2 \int_0^\infty dE \rho(E) E \Bigg(1 - \frac{1}{e^{-\frac{E}{k_B T}} + 1}\Bigg) \\
&\quad + 2 \int_0^\infty dE \rho(E) E \frac{1}{e^{\frac{E}{k_B T}} + 1} \\
&= 2 \int_0^\infty dE \rho(E) E \frac{e^{-\frac{E}{k_B T}} + 1}{\cosh{\Big(\frac{E}{k_B T}\Big)} + 1}.
\end{split}
\end{align}
The density of states at band energy $E$ can be solved by applying the dispersion relationship as follows:
\begin{align} \label{eq: electron density of states}
\begin{split}
\rho(E) &= \frac{dN}{dA_k} \frac{dA_k}{dk} \frac{dk}{dE} \\
&= \frac{A}{(2\pi)^2} \bigg(2\pi \frac{E}{\hbar v_F}\bigg) \frac{1}{\hbar v_F} \\
&= \frac{A}{2\pi \hbar^2 v_F^2} E,
\end{split}
\end{align}
where $A$ is the surface state area, $A_k$ is the area in the reciprocal space associated with the value $k = |\bm{k}|$, and $N$ is the number of states. As expected for the graphene-like band structure, the density of states is linear in the energy. We are thus in a position to solve for the electron energy $U_{el}(T)$ as a function of temperature $T$:
\begin{align}
\begin{split}
U_{el}(T) &= \frac{A}{\pi \hbar^2 v_F^2} \int_0^\infty dE E^2 \frac{e^{-\frac{E}{k_B T}} + 1}{\cosh{\Big(\frac{E}{k_B T}\Big)} + 1} \\
&= \frac{A}{\pi \hbar^2 v_F^2} 3 (k_B T)^3 \zeta(3),
\end{split}
\end{align}
where $\zeta$ represents the Riemann zeta function, with $\zeta(3) \approx 1.2$. This implies that the electron temperature varies with the total electron energy as $U_{el}^{1/3}$, and the relationship between the gain in electron energy and the temperature change from $T_i$ to $T_f$ takes the following form:
\begin{equation} \label{eq: electron energy gain}
\Delta U_{el} = \frac{3.6 A k_B^3}{\pi \hbar^2 v_F^2} \Big(T_f^3 - T_i^3\Big).
\end{equation}
For detecting a moderate number of microwave photons, we are primarily interested in the limit $\Delta T (\equiv T_f - T_i) \ll T_i$ where $T_i > 0.1 \textrm{ K}$. In that limit, the cooling power $Q$ for the electron modes is related to the rate of change of the temperature by taking the derivative of $\Delta U_{el}$ with respect to time, yielding the following function of the temperature $T \approx T_i$:
\begin{equation} \label{eq: surface electron cooling power}
	Q = -\frac{d}{dt} \bigg(\frac{3.6 A k_B^3}{\pi \hbar^2 v_F^2} T^3\bigg) = -\frac{10.8 A k_B^3}{\pi \hbar^2 v_F^2} T^2 \frac{dT}{dt}.
\end{equation}

\subsection{Energy Gain for Bulk Phonon Modes}
Next, we look to determine how the energy gained by the lattice vibrations relates to the lattice temperature. In general, the total phonon energy is determined as a function of temperature by summing over the modes corresponding to various phonon branches $\mu$ and phonon wavevectors $\bm{q}$, weighted by the occupation number $\expect{n_{\mu,\bm{q}}}$ for each phonon mode:
\begin{equation} \label{eq: total phonon energy summation}
	U_{ph}(T) = \sum_\mu \sum_{\bm{q}} \expect{n_{\mu,\bm{q}}(T)} \hbar \omega_{\mu,\bm{q}},
\end{equation}
where $\omega_{\mu,\bm{q}}$ is the frequency of the phonon mode of branch $\mu$ and wavevector $\bm{q}$, and the occupation number at a given temperature $T$ is calculated from the Bose-Einstein distribution:
\begin{equation}
\expect{n_{\mu,\bm{q}}(T)} = \frac{1}{e^{\frac{\hbar \omega_{\mu,\bm{q}}}{k_B T}} - 1}.
\end{equation}
As previously mentioned, the fact that the sample is in the low-temperature regime (below 0.5 K) indicates that the Debye model, with its assumption of a linear phonon dispersion, is approximately valid for the phonon modes with non-negligible occupation numbers. Therefore, we can restrict the summation over branches to just the 3 acoustic branches (corresponding to the 3 polarizations). In general, the slope of each of these branches is slightly anisotropic in reciprocal space due to the varying angle between the polarization and propagation directions. However, per the treatment in Kittel \cite{Kittel}, we can approximate the composite effect of the 3 branches on the summation in Eq.~\eqref{eq: total phonon energy summation} as equivalent to a summation over 3 isotropic branches, each featuring a slope of $v_s$ (i.e., the speed of sound in the material), such that:
\begin{align}
\begin{split}
U_{ph}(T) = 3 \sum_{\bm{q}} \frac{\hbar \omega_q}{e^{\frac{\hbar \omega_q}{k_B T}} - 1},
\end{split}
\end{align}
where $\omega_q = v_s q$, and the speed of sound $v_s$ for Cd\textsubscript{3}As\textsubscript{2} is estimated as $2.3 \times 10^3 \textrm{ m/s}$ \cite{LundgrenFiete}.
\par 
We now replace the summation over wavevectors with an integral over the density of states for each branch in terms of frequency, $D(\omega)$. For a 3-dimensional lattice with a speed of sound $v_s$, this density is determined as follows:
\begin{align}
\begin{split}
D(\omega) &= \frac{dN}{dV_q} \frac{dV_q}{dq} \frac{dq}{d\omega} \\
&= \frac{V}{(2\pi)^3} \Bigg(4 \pi \bigg(\frac{\omega}{v_s}\bigg)^2\Bigg) \frac{1}{v_s} \\
&= \frac{V \omega^2}{2\pi^2 v_s^3},
\end{split}
\end{align}
where $V$ is the bulk volume of the lattice and $V_q$ is the volume in the reciprocal space associated with $q = |\bm{q}|$. We therefore solve for the total phonon energy as a function of temperature through the following integral:
\begin{align}
\begin{split}
U_{ph}(T) &= 3 \int_0^\infty d\omega D(\omega) \frac{\hbar \omega}{e^{\frac{\hbar \omega}{k_B T}} - 1} \\
&= \frac{3V \hbar}{2\pi^2 v_s^3} \int_0^\infty d\omega \frac{\omega^3}{e^{\frac{\hbar \omega}{k_B T}} - 1}.
\end{split}
\end{align}
Note that the expression differs from that in Kittel in that we use $\infty$ instead of a specific Debye cutoff for the upper bound of the frequency range. As in the case of the integral over energy for the electronic modes in the previous section, this is justified by the rapid convergence of the integrand to 0 at very low temperatures \cite{DebyeLibretexts}, corresponding to the fact that only the linear regime of the acoustic branches are non-negligibly occupied. Then, we find the following result:
\begin{align}
\begin{split}
U_{ph}(T) &= \frac{3V k_B^4 T^4}{2\pi^2 v_s^3 \hbar^3} \int_0^\infty d\bigg(\frac{\hbar \omega}{k_B T}\bigg) \frac{\Big(\frac{\hbar \omega}{k_B T}\Big)^3}{e^{\frac{\hbar \omega}{k_B T}} - 1} \\
&= \bigg(\frac{3V k_B^4 T^4}{2\pi^2 v_s^3 \hbar^3}\bigg) \bigg(\frac{\pi^4}{15}\bigg)
\end{split}
\end{align}
Unlike the total surface electron energy, which scales as $T^3$, the total bulk phonon energy scales as $T^4$. This difference can be attributed to the fact that the surface is 2D whereas the bulk is 3D, having more degrees of freedom, thus implying that all else being equal, a given change in energy would have a weaker effect on bulk temperature than on surface temperature.
\par 
The change in the total phonon energy can therefore be related to the initial and final temperatures $T_i$ and $T_f$ as follows:
\begin{equation} \label{eq: phonon energy gain}
\Delta U_{ph} = \frac{\pi^2 V k_B^4}{10 \hbar^3 v_s^3} \Big(T_f^4 - T_i^4\Big).
\end{equation}
Note that, for a prism-shaped sample (such as a thin film) for which the surface state forms one of the two bases for the prism, the bulk volume is proportional to the surface area as $V = Ad$, where $d$ represents the sample thickness.

\subsection{Comparison of Specific Heat}
\begin{figure}[!tb]
	\centering
	\includegraphics[width=\linewidth]{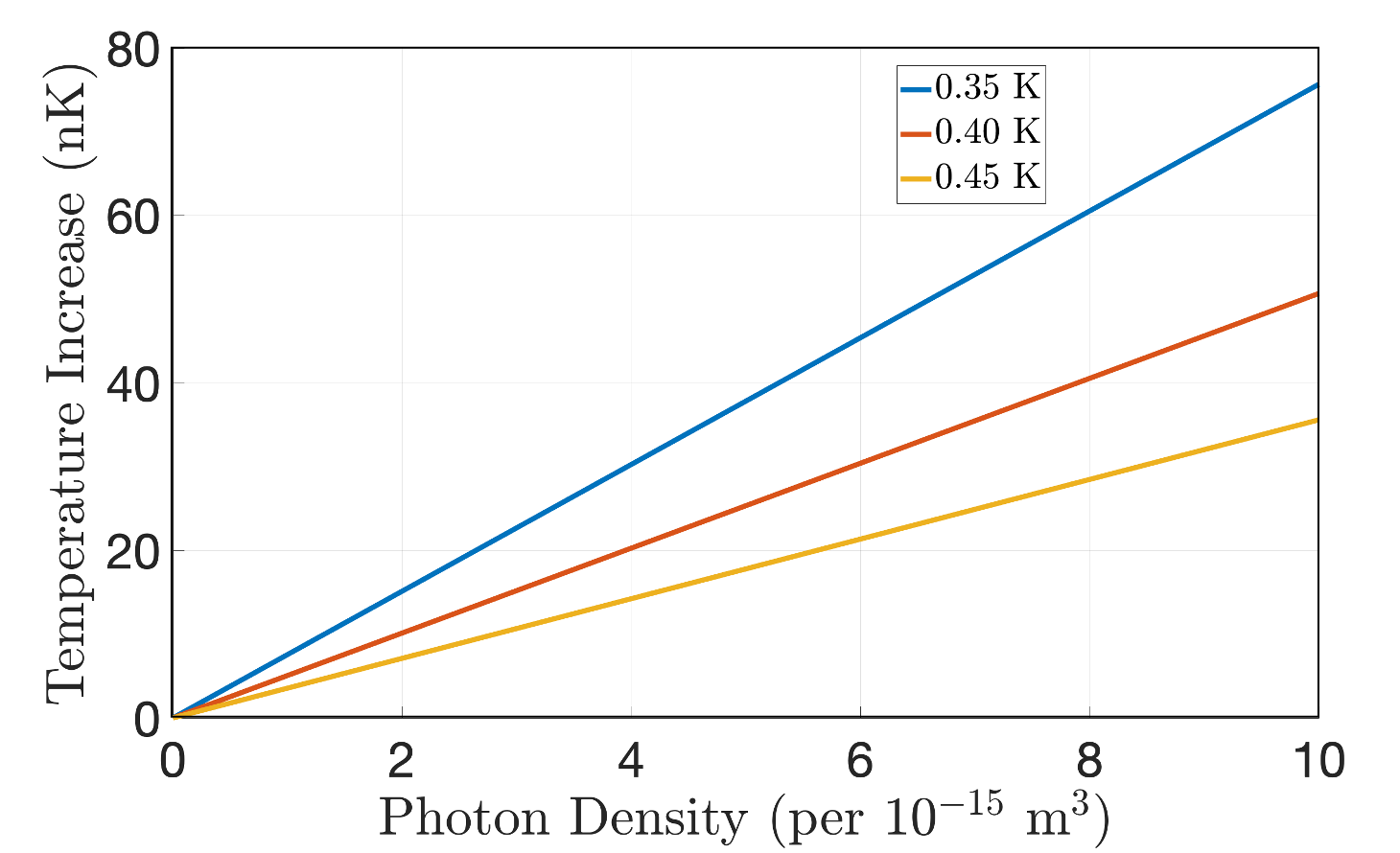}
	\caption{Plots of temperature gain (in nanokelvins) versus density of absorbed photons (per $10^{-15} \textrm{ m}^{3}$) for baseline temperatures $T_i = 0.35, 0.40, \textrm{ and } 0.45 \textrm{ K}$ given a photon frequency $f = 5 \textrm{ GHz}$.}
	\label{fig:temperaturegraph}
\end{figure}

Having derived the energy gain for the surface electron and the bulk phonon modes for given initial and final temperatures, we now seek to compare the specific heat values for the two mode types in order to glean an understanding of how the excess thermal energy is distributed between the modes. From the result in Eq.~\eqref{eq: electron energy gain}, the specific heat for the collection of surface electrons is determined as follows:
\begin{equation} \label{eq: specific heat for electrons}
	C_{el}(T) = \frac{dU_{el}}{dT} = \frac{10.8 A k_B^3}{\pi \hbar^2 v_F^2} T^2.
\end{equation}
The specific heat for the bulk lattice is calculated in an analogous manner from the result in Eq.~\eqref{eq: phonon energy gain}:
\begin{equation} \label{eq: specific heat for phonons}
	C_{ph}(T) = \frac{dU_{ph}}{dT} = \frac{2\pi^2 V k_B^4}{5 \hbar^3 v_s^3} T^3.
\end{equation}
Using the relationship $V = Ad$, where $d$ denotes the bulk depth of the lattice, we divide Eq.~\eqref{eq: specific heat for phonons} by~\eqref{eq: specific heat for electrons} in order to determine the ratio of the specific heat values for a given temperature:
\begin{align}
\begin{split}
\frac{C_{ph}(T)}{C_{el}(T)} &= \frac{\pi^3 k_B v_F^2}{27 \hbar v_s^3} T d \\
&= \Big(1.24 \times 10^{13} \textrm{ K}^{-1}\textrm{m}^{-1}\Big) T d.
\end{split}
\end{align}
Since the baseline refrigerator temperature is at least 0.25 K, this sets the minimum value for $T$. The lattice depth $d$ must be at least multiple times longer than the lattice constant, which is $3-5 \textrm{ \AA}$ for Cd\textsubscript{3}As\textsubscript{2} \cite{WangLatticeStructure}. Therefore, the phonon specific heat is far higher than the electron specific heat (by at least 3 orders of magnitude), indicating that nearly all of the thermal energy gained from the photon absorption is eventually stored in the lattice vibrational modes. As such, if $N$ photons of frequency $\omega$ are absorbed, then the relationship between the final equilibrium temperature $T_f$ and the initial temperature $T_i$ can be determined from Eq.~\eqref{eq: phonon energy gain}:
\begin{align} \label{eq: final temperature}
\begin{split}
T_f &= \Bigg(T_i^4 + \frac{10\hbar^3 v_s^3}{\pi^2 V k_B^4} N\hbar\omega\Bigg)^{1/4} \\
&= \Bigg(T_i^4 + \Big(4.13 \times 10^{-35} \textrm{ m}^3\textrm{K}^4\textrm{s}\Big) \frac{N\omega}{V}\Bigg)^{1/4}
\end{split}
\end{align}
As this expression shows, the determinitive factor in the temperature increase is the total photon energy absorbed per unit volume of the lattice, which is proportional to $N\omega/V$. We plot the temperature gain as a function of photoelectron density $N/V$ for 3 different baseline temperatures 0.35 K, 0.40 K, and 0.45 K, for an input photon frequency of 5 GHz (corresponding to $\omega = \pi \times 10^{10} \textrm{ s}^{-1}$) in Fig.~\ref{fig:temperaturegraph}.
Note that for low photon densities such that the temperature gain is small compared to the baseline temperature, the relationship between temperature gain and photon density is approximately linear, as expected.
\par
The imbalance between the electron and lattice specific heat values also has significant implications for the heat transfer between the electron and phonon distributions that ultimately yields the equilibrium state. In particular, when comparing the quasiequilibrium electron and lattice temperatures to the final equilibrium temperature for the whole system, the equilibrium temperature will be much closer to the quasiequilbrium lattice temperature than to the electron temperature. We consider the timescale for the electron-lattice heat transfer as a function of temperature in the next section.

\section{Electron-Phonon Interaction Timescale} \label{sec: Electron-Phonon Interaction Timescale}

Having derived the equilibrium temperature for the bulk lattice upon photon absorption, we now aim to estimate the timescale over which that equilibrium is reached. As previously mentioned, this energy transfer takes place in two steps: a rapid electron-electron rethermalization, followed by heat transfer from the electrons to the acoustic vibrations of the lattice (which is much slower than the electron-electron interaction \cite{UltrafastDiracFermionRelaxation,GrapheneUltrafastCarrierDynamics,LundgrenFiete}). Here, we will focus on the latter process, since it serves as the limiting factor in setting the minimum timescale for reaching equilibrium. The heat transfer timescale between bulk electrons and lattice phonons in Cd\textsubscript{3}As\textsubscript{2} has been the subject of recent analysis \cite{LundgrenFiete,KubakaddiBiswas}, and here we will build on that analysis to solve for the heat transfer timescale between surface electrons and lattice phonons. We will characterize the available phase space area for bulk phonon emission by the surface electrons, derive the matrix element for the electron-phonon interaction, and finally calculate the rate for the electron-phonon heat transfer.

\subsection{Phase Space}
\begin{figure*}[!tb]
	\centering
	\begin{subfigure}{\columnwidth}
		\centering
		\includegraphics[width=0.5\linewidth]{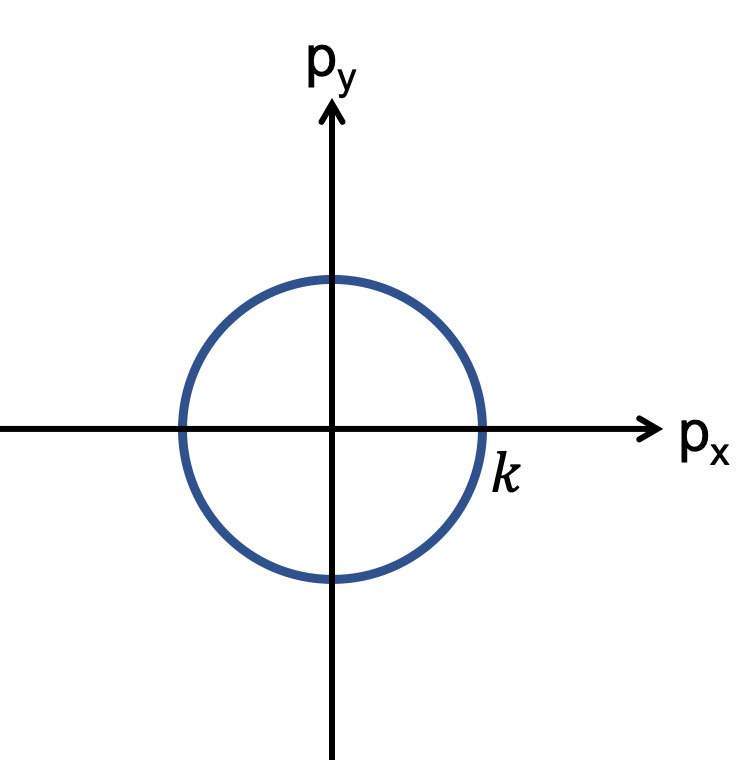}
		\caption{}
	\end{subfigure}
	\begin{subfigure}{\columnwidth}
		\centering
		\includegraphics[width=\linewidth]{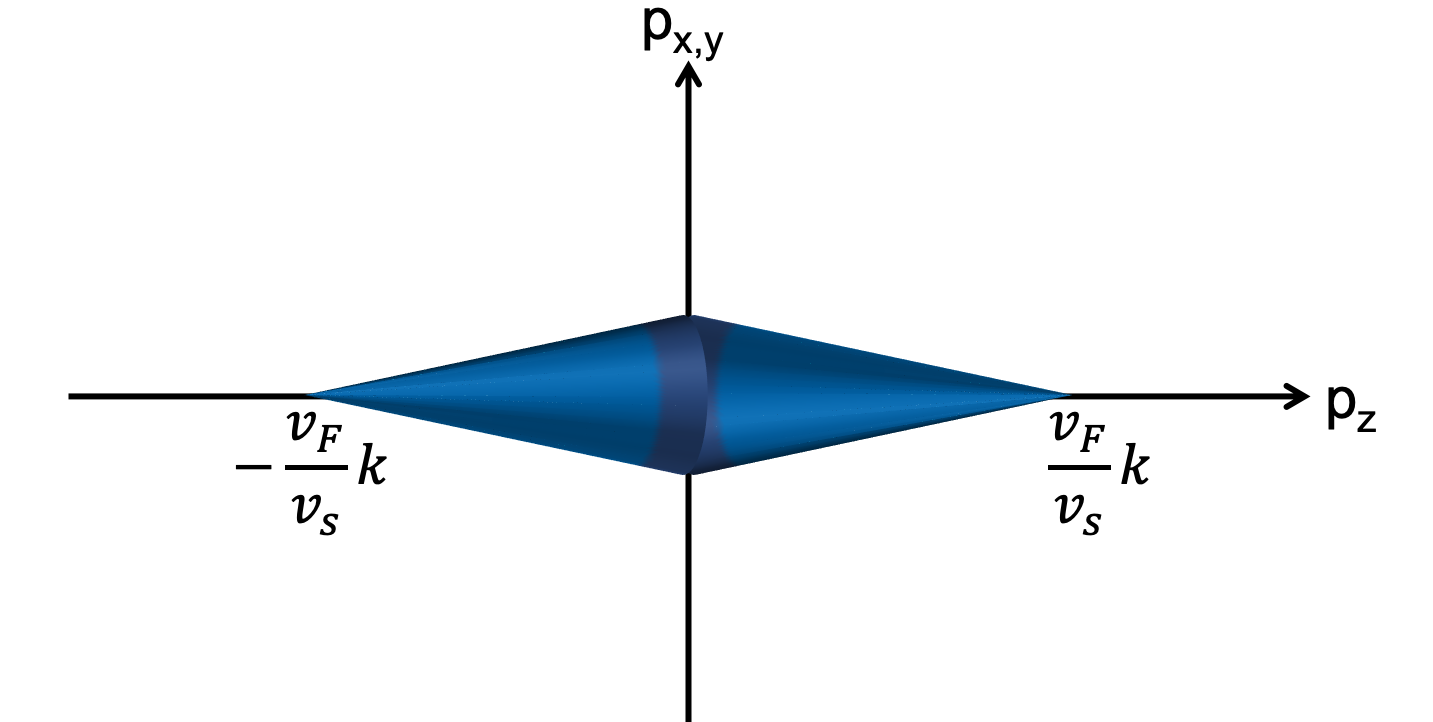}
		\caption{}
	\end{subfigure}
	\caption{Depiction of the phase space area of final electronic states for a given initial state $\bm{k} = (k,0,0)$, including the 2D cross-section of the phase space along the $p_xp_y$-plane (a) and the phase space along the $p_z$-axis (b). The phase space forms a double cone, with a base of radius $k$ on the $p_xp_y$-plane, and tapering off in the $+p_z$ and $-p_z$ directions with a length of $v_Fk/v_s$ along each. Note that the base versus height ratio for the double-cone in (b) is not to scale.}
	\label{fig:phasespace}
\end{figure*}

We start by examining the available phase space for the interaction between 2D surface state electrons and the bulk phonon modes. Unlike the spherical equal-energy manifolds for the 3D Dirac cone carrier modes, the 2D Dirac cone electron modes take a cylindrical equal-energy manifolds, with a degree of freedom in the $k_z$-direction (corresponding to the axis perpendicular to the surface). We therefore use cylindrical coordinates, expanding the final electron wavevector $\bm{p}$ as $(p\cos\theta_{\bm{p}},p\sin\theta_{\bm{p}},p_z)$ and the initial wavevector $\bm{k}$ as $(k,0,0)$. In this coordinate system the emitted phonon wavevector $\bm{q}$ can be expanded as follows:
\begin{equation} \label{eq: phonon coordinates}
	\bm{q} = \bm{k - p} = (p\cos\theta_{\bm{p}} - k, p\sin\theta_{\bm{p}}, p_z).
\end{equation}
Note that the bulk phonon's equal-energy manifolds retain a 3D spherical shape defined by $\omega = v_s q$. We map this onto the electron's manifolds using energy conservation:
\begin{equation}
v_F k - v_F p = v_s \sqrt{(p\cos{\theta_{\bm{p}}} - k)^2 + p^2\sin^2{\theta_{\bm{p}}} + p_z^2}.
\end{equation}
Squaring both sides and solving for $p$, we find that $p$ varies with both $\theta_{\bm{p}}$ and $p_z$:
\begin{align}
\begin{split}
p &= k\Bigg(1 + \Delta(\theta_{\bm{p}}) \\
&\quad - \sqrt{\Big(1 + \Delta(\theta_{\bm{p}})\Big)^2 - 1 + \bigg(\frac{v_s^2}{v_F^2 - v_s^2}\bigg) \frac{p_z^2}{k^2}}\Bigg),
\end{split}
\end{align}
where $\Delta(\theta_{\bm{p}})$ is defined as follows:
\begin{equation} \label{eq: Delta(theta)}
\Delta(\theta_{\bm{p}}) = \frac{v_F^2 - v_s^2\cos{\theta_{\bm{p}}}}{v_F^2 - v_s^2} - 1.
\end{equation}
For our material, $v_s \ll v_F$, thus yielding $\Delta(\theta_{\bm{p}}) \ll 1$ for all $\theta_{\bm{p}}$. Therefore, $p$ can be approximately expressed as solely a linear function of $p_z$:
\begin{equation}
p \approx k - \frac{v_s}{v_F} |p_z|.
\end{equation}
Geometrically, the available phase space can be envisioned as pair of cones aligned along the $p_z$-axis, with the bases overlapping at $p_z = 0$, as depicted in Fig.~\ref{fig:phasespace}.
The radius attains is maximal value of $k$ at $p_z = 0$ and tapers off as the magnitude of $p_z$ increases. The vertices are reached at the following values of $p_z$:
\begin{equation} \label{eq: p_z max}
|p_z|_{max} \approx \frac{v_F}{v_s} k.
\end{equation}
Finally, we determine the phonon wavevector $q$ from the calculated value of $\bm{p}$ as a function of $\bm{k}$. As shown in Eq.~\eqref{eq: phonon coordinates}, the amplitudes of $p_z$ and $q_z$ will equal each other, since the electron and phonon dispersion centers lie on the same $xy$-plane. We label $q_{xy}$ as the component of $\bm{q}$ perpendicular to the $q_z$-axis. For a given $p_z$ and $\theta_{\bm{p}}$ pair, the amplitudes $q_{xy}$, $k$, and $p$ form the 3 legs of a triangle for which $\theta_{\bm{p}}$ represents the angle between the sides of lengths $k$ and $p$. Therefore, $q_{xy}$ can be calculated as follows:
\begin{equation}
q_{xy} \approx \sqrt{k^2 + K^2 - 2kK\cos{\theta_{\bm{k}}}},
\end{equation}
and since $q_z = -p_z$, the frequency $\omega_q$ of the emitted phonon is straightforwardly calculated from the speed of sound:
\begin{align}
\begin{split}
\omega_q &= v_s \sqrt{q_{xy}^2 + q_z^2} \\
&\approx v_s \sqrt{p^2 + k^2 - 2pk\cos{\theta_{\bm{p}}} + p_z^2}.
\end{split}
\end{align}
Since $v_F \gg v_s$ (by a factor of 400), the length of each cone is far longer than the diameter, implying that the approximation $q \approx |q_z| = |p_z| \approx p$ will be valid for nearly all of the available phase space. 

\subsection{Heat Transfer Rate}
Having determined the phase space for the electron-phonon interaction, we are now in a position to calculate the heat transfer rate between the two modes from the composite interaction. Labeling the electron energy for a generic wavevector $\bm{k'}$ as $E_{\bm{k'}}$, the matrix element corresponding to the electronic transition from $\bm{k}$ to $\bm{p}$ through the emission of a phonon in branch $\mu$ and wavevector $\bm{q}$ as $M_{\bm{k},\bm{p}}^{\mu,\bm{q}}$, the Bose-Einstein phonon occupation number for the mode frequency $\omega$ at temperature $T$ as $n_T(\omega)$, and the Fermi-Dirac distribution value at $T$ as $f(T)$, the rate $Q$ is determined through the following summation over initial carrier wavevectors $\bm{k}$, final carrier wavevectors $\bm{p}$, and phonon branches and wavevectors $(\mu,\bm{q})$ \cite{ViljasGrapheneHeatTransfer, HotElectrons}:
\begin{align}
\begin{split}
Q &= \frac{2\pi}{\hbar} \sum_{\bm{k}} \sum_{\bm{p}} \sum_{\mu,\bm{q}} \Big(E_{\bm{k}} - E_{\bm{p}}\Big) \Big|M_{\bm{k},\bm{p}}^{\mu,\bm{q}}\Big|^2 \\
&\quad \times \Big(f(E_{\bm{k}}) - f(E_{\bm{p}})\Big) \Big(n_{T_L}(\omega_{\mu,\bm{q}}) - n_{T_e}(\omega_{\mu,\bm{q}})\Big) \delta_{\bm{k},\bm{p+q}} \\
&\quad \times \delta\Big(E_{\bm{k}} - E_{\bm{p}} - \hbar \omega_{\mu,\bm{q}}\Big).
\end{split}
\end{align}
As previously discussed, the low temperature restricts the occupied phonon modes to the long-wavelength acoustic regime. The interaction between electrons and long-wavelength acoustic phonons is dominated by the deformation potential \cite{BardeenShockleyDP, HerringVogtDP}, as recently applied to the interaction between bulk electrons and phonons in Cd\textsubscript{3}As\textsubscript{2} \cite{LundgrenFiete, BhargaviKubakaddi}. As discussed in Appendix~\ref{sec: Matrix Elements for Interaction Between Surface Electrons and Bulk Phonons}, the equivalent matrix elements apply for the interaction between surface electrons and bulk phonons. Therefore, the matrix element amplitude-squared reduces to a function varying solely with and linear in the phonon amplitude $q$:
\begin{equation} \label{eq: bulk matrix element amplitude-squared}
\sum_{\mu} \Big|M_{\bm{k},\bm{p}}^{\mu,\bm{q}}\Big|^2 = \frac{C}{V} q,
\end{equation}
where $V$ represents the lattice volume and $C$ is a constant that varies with the square of the deformation potential. Substituting this, along with the electron and phonon dispersion relationships into the expression for $Q$, we find that it takes the following form:
\begin{align}
\begin{split}
Q &= \frac{2\pi}{\hbar} \sum_{\bm{k},\bm{p},\bm{q}} \hbar v_F \Big(k - p_{xy}\Big) \frac{C}{V} q \Big(f(\hbar v_F k) - f(\hbar v_F p_{xy})\Big) \\
&\quad \times \Big(n_{T_L}(v_s q) - n_{T_e}(v_s q)\Big) \\
&\quad \times \delta_{\bm{k},\bm{p+q}} \delta\Big(\hbar v_F k - \hbar v_F p_{xy} - \hbar v_s q\Big).
\end{split}
\end{align}
The summation is simplified in Appendix~\ref{sec: Calculating the Surface Electron Cooling Rate} in the limit $\Delta T \ll T$, where $T \approx T_e \approx T_L$ and $\Delta T$ is defined as $T_e - T_L$. We find that the Dirac and Kronecker delta functions combine to reduce the integral over the phase space volume to the double-cone phase space area derived previously, as expected. This yields the following expression for the carrier-phonon heat transfer rate due to intraband (valence-valence or conduction-conduction) transitions:
\begin{widetext}
\begin{align}
	\begin{split}
	Q &\approx \frac{AC}{2\pi^2 \hbar} \frac{v_s}{v_F} \bigg(\frac{\hbar v_s}{k_B T}\bigg) \frac{\Delta T}{T} \bigg(\frac{k_B T}{\hbar v_F}\bigg)^3 \bigg(\frac{k_B T}{\hbar v_s}\bigg)^4 \int_0^{\infty} dx x \int_0^x dy (x-y) y^3 \frac{e^y}{(e^y - 1)^2} \Bigg(\frac{1}{e^{x} + 1} - \frac{1}{e^{x-y} + 1}\Bigg).
	\end{split}
\end{align}
\end{widetext}
Solving the integral numerically, we obtain a value of $-32$. Therefore, $Q$ is further reduced to the following:
\begin{equation}
Q \approx -\frac{16 A C k_B^6}{\pi^2 \hbar^7 v_F^4 v_s^2} T^5 \Delta T.
\end{equation}
Next, we solve for the heat transfer rate due to interband transitions. Based again on Appendix~\ref{sec: Calculating the Surface Electron Cooling Rate}, we use the following expression:
\begin{widetext}
	\begin{align}
	\begin{split}
	Q_{inter} &\approx \frac{AC}{2\pi^2 \hbar} \frac{v_s}{v_F} \bigg(\frac{\hbar v_s}{k_B T}\bigg) \frac{\Delta T}{T} \bigg(\frac{k_B T}{\hbar v_F}\bigg)^3 \bigg(\frac{k_B T}{\hbar v_s}\bigg)^4 \int_0^{\infty} dx x \int_x^{\infty} dy (y-x) y^3 \frac{e^y}{(e^y - 1)^2} \Bigg(\frac{1}{e^{x} + 1} - \frac{1}{e^{x-y} + 1}\Bigg).
	\end{split}
	\end{align}
\end{widetext}
Note that the constants in front of the integral are identical to that for the intraband case. Solving this integral numerically yields a value of $-100$. The total heat transfer rate $Q_{total}$ from the surface carriers to the lattice vibrations is determined by multiplying the intraband rate $Q$ by 2 (to account for both bands) and then summing with the interband rate $Q_{inter}$:
\begin{align}
\begin{split}
Q_{total} &= 2Q + Q_{inter} \\
&\approx -\frac{82 A C k_B^6}{\pi^2 \hbar^7 v_F^4 v_s^2} T^5 \Delta T.
\end{split}
\end{align}
In order to determine the heat transfer timescale, we substitute the previously derived relationship between the electron cooling rate and the rate of change of electron temperature from Eq.~\eqref{eq: surface electron cooling power} into the left-hand-side of the above expression:
\begin{align}
\begin{split}
-\frac{10.8 A k_B^3}{\pi \hbar^2 v_F^2} T^2 \frac{d(\Delta T)}{dt} &\approx -\frac{82 A C k_B^6}{\pi^2 \hbar^7 v_F^4 v_s^2} T^5 \Delta T, \\
\frac{d(\Delta T)}{dt} &\approx -\frac{7.6 C k_B^3 T^3}{\pi \hbar^5 v_F^2 v_s^2} \Delta T.
\end{split}
\end{align}
As the result shows, the electron temperature decays exponentially toward the lattice temperature, with the rate varying as $T^3$. 
\par 
The remaining task is to determine the value of the constant $C$, which derives from the electron-phonon matrix element. One method for doing so is by using the deformation potential of 20 eV measured by Jay-Gerin \textit{et al.} \cite{DeformationPotential}. This yields the following value for $C$, using a material density $\rho = 7 \times 10^3 \textrm{ kg/m}^3$ \cite{LundgrenFiete}:
\begin{equation}
	C = \frac{\hbar D^2}{4 \rho v_s} = 1.7 \times 10^{-77} \textrm{ J}^2\textrm{m}^4.
\end{equation}
This leads to the following heat transfer time constant $\gamma$:
\begin{equation}
	\gamma \approx \frac{7.6 C k_B^3 T^3}{\pi \hbar^5 v_F^2 v_s^2} \approx \Big(1.6 \times 10^6 \textrm{ K}^{-3}\textrm{s}^{-1}\Big) T^3.
\end{equation}
An alternative method for finding the deformation potential is by merging the experimental results from Weber \textit{et al.} \cite{WeberAcousticPhononEmission} with the theory provided by Lundgren and Fiete \cite{LundgrenFiete}. Specifically, Weber \textit{et al.} used a bulk Cd\textsubscript{3}As\textsubscript{2} sample intrinsically doped to a baseline electron density of $6 \times 10^{23} \textrm{ m}^{-3}$, which corresponds to a Fermi energy of 170 meV and a Fermi temperature of 1130 K. Under these conditions, they observed a timescale of 3.1 ps for electron cooling by low-energy acoustic phonon emission at lattice temperatures of 80 K and 300 K. This scenario is addressed by Lundgren and Fiete's Equation (8), which models the heat transfer rate for $k_B T \ll E_f $ (where $E_f$ is the Fermi energy):
\begin{equation} \label{eq: Equation 8 Lundgren Fiete}
	\gamma = \frac{D^2 E_f^4}{3 k_B \hbar^4 v_F^5 \rho T}.
\end{equation}
We now substitute a temperature and rate data point from Weber \textit{et al.} into this expression to calculate the deformation potential $D$. Since the limit $k_B T \ll E_f$ is much more valid for $T = 80 \textrm{ K}$ than for 300 K, we use the former as the temperature corresponding to the rate $\gamma = (3.1 \textrm{ ps})^{-1}$ for the purposes of application to Eq.~\eqref{eq: Equation 8 Lundgren Fiete}. This yields the following value for $D$:
\begin{equation}
	D = \bigg(\frac{3 k_B \hbar^4 v_F^5 \rho}{E_f^4} T \gamma\bigg)^{\frac{1}{2}} = 250 \textrm{ eV}.
\end{equation}
This leads to the following value for the coefficient $C$:
\begin{equation}
	C = \frac{\hbar D^2}{4 \rho v_s} = 2.7 \times 10^{-75} \textrm{ J}^2\textrm{m}^4,
\end{equation}
which yields the following heat transfer time for our model:
\begin{equation}
	\gamma \approx \frac{7.6 C k_B^3 T^3}{\pi \hbar^5 v_F^2 v_s^2} \approx \Big(2.5 \times 10^8 \textrm{ K}^{-3}\textrm{s}^{-1}\Big) T^3.
\end{equation}
As will be discussed in the next section, the lower bound for the baseline temperature $T$ (which will also set the minimum value for the heat transfer rate) will be about 0.35 K. For this temperature, the above two methods yield a lattice heating timescale approximately ranging from 93 ns to 15 $\mu$s. 

It is worth comparing this timescale with the corresponding timescale for heat transfer between lattice phonons and bulk electrons (when the bulk is in the normal, non-superconducting phase). Based on Eq. (6) of Lundgren and Fiete \cite{LundgrenFiete}, this timescale would be on the order of 7000 seconds, well over 8 orders of magnitude longer than even the upper bound value for the transfer time from surface electrons to the lattice phonons. This difference can be attributed to the vastly greater available phase space area for the surface electron interaction. Consequently, any heat transfer from the phonons to the bulk electrons is insignificant compared to that from the surface electrons to the phonons. 

\section{Photon-number resolving detection} \label{sec: Discussion}
We now describe the photon-number resolving detector scheme based on our theoretical findings. First, we address the question of whether the timescale for lattice temperature equilibration is much faster than the dissipation time through thermal conduction or radiative decay. Regarding the thermal conduction heat loss, we note that the contacts used for cooling the sample can be removed after the material reaches the refrigerator temperature. As a result, the heat dissipation time through thermal conduction will range on the order of several hours and can thus be ignored. Instead, we will focus on the radiative loss. Based on the results calculated for graphene, the electron-hole interband dipole moment for a 2D Dirac cone band structure is given as a function of photon radial frequency $\omega$ as follows \cite{GrapheneDipole}:
\begin{equation}
d_{c,v} = \frac{ev_F}{\omega}.
\end{equation}
Substituting this into the well-known radiative decay rate expression based on the Einstein coefficients \cite{EinsteinCoefficients}, we find that the radiative rate varies linearly with $\omega$:
\begin{align}
\begin{split}
\Gamma_{rad}(\omega) &= \frac{\omega^3}{3\pi \epsilon_0 \hbar c^3} \Big|d_{c,v}(\omega)\Big|^2 \\
&= \frac{e^2 v_F^2}{3\pi \epsilon_0 \hbar c^3} \omega \\
&= \Big(1.1 \times 10^{-7}\Big) \omega
\end{split}
\end{align}
For frequencies up to 10 GHz, the radiative decay time is therefore 150 $\mu$s or greater. This is significantly longer than the electron-phonon heat transfer time calculated above, which is 15 $\mu$s or less, which in turn is much longer than the previously discussed electron-electron rethermalization time of tens of picoseconds \cite{MihnevCarrierRelaxationTime}. Therefore, a rapid rethermalization of the electron population in the bands occurs before any radiative loss of the photoelectrons occurs.

Next, we address the question of heat transfer from the surface electronic modes directly to the bulk electronic modes. This would constitute a loss process, since it reduces the heat absorbed by the bulk phonon modes. We note that the aforementioned spatial separation between bulk and surface electronic states renders this process unlikely. It is also worth comparing the heat capacity of the bulk electron modes to that of the phonon modes. To this end, in the temperature range 0.35-0.45 K (just over $0.5T_c$), the superconducting state features approximately the same heat capacity as the normal state extrapolated to that temperature range. As such, we use the collective electron energy expression shown in Eq.~\eqref{eq: electron energy increase integral}, this time using the 3D rather than 2D Dirac cone dispersion to derive the density of states $\rho(E)$:
\begin{align}
\begin{split}
\rho(E) &= \frac{dN}{dV_k} \frac{dV_k}{dk} \frac{dk}{dE} \\
&= \frac{V}{(2\pi)^3} \Bigg(4\pi \bigg(\frac{E}{\hbar v_F}\bigg)^2\Bigg) \frac{1}{\hbar v_F} \\
&= \frac{V}{2\pi^2 \hbar^3 v_F^3} E^2,
\end{split}
\end{align}
where $V$ denotes the bulk volume. Substituting into Eq.~\eqref{eq: electron energy increase integral}, we find the following bulk thermal energy as a function of temperature:
\begin{align}
\begin{split}
U_{el,bulk}(T) &= \frac{V}{\pi^2 \hbar^3 v_F^3} \int_0^\infty dE E^3 \frac{e^{-\frac{E}{k_B T}} + 1}{\cosh{\Big(\frac{E}{k_B T}\Big)} + 1} \\
&= \frac{V}{\pi^2 \hbar^3 v_F^3} (k_B T)^4 \frac{7\pi^4}{60} \\
&= \frac{7\pi^2 V k_B^4}{60 \hbar^3 v_F^3} T^4.
\end{split}
\end{align}
The bulk heat capacity is calculated by taking the derivative with respect to the temperature $T$:
\begin{equation}
C_{el,bulk}(T) = \frac{dU_{el,bulk}}{dT} = \frac{7\pi^2 V k_B^4}{15 \hbar^3 v_F^3} T^3.
\end{equation}
Comparing this to the phonon heat capacity (see Eq.~\eqref{eq: specific heat for phonons}), we find that the phonon heat capacity is greater by a factor of approximately $(v_F/v_s)^3$, i.e. more than 7 orders of magnitude. This massive disparity can be explained by the fact that near the Fermi level, the electron group velocity vastly exceeds the phonon group velocity, resulting in a far greater density of states for the phonon modes than for the electron modes. We thus conclude that the energy of the absorbed photons is safely transferred to, first, the rethermalization of the carrier band populations, and then, to the bulk phonon modes to elevate the bulk temperature.

\begin{figure}[!tb]
	\centering
	\includegraphics[width=\linewidth]{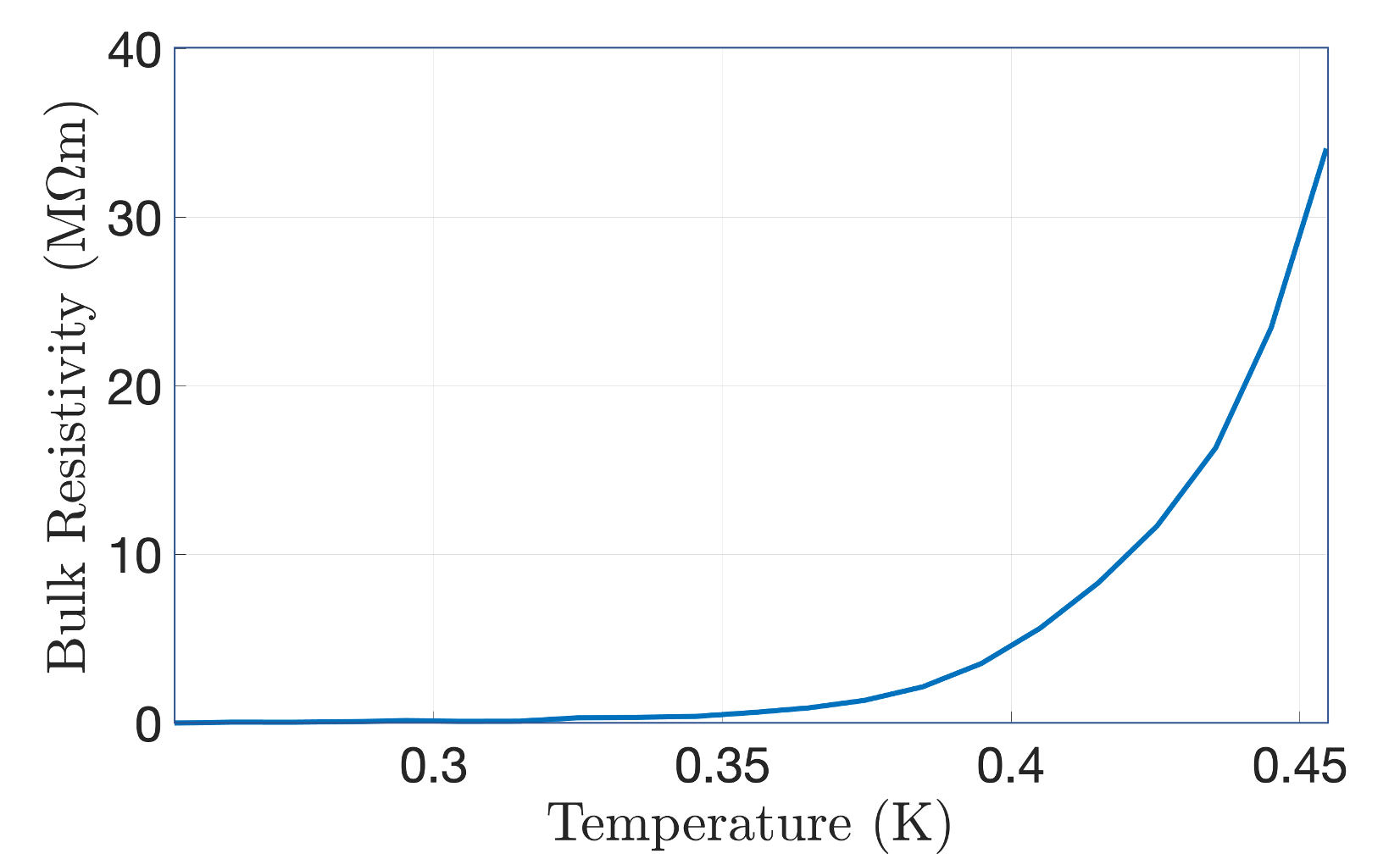}
	\caption{Resistivity (in megaohm-meters) versus temperature (in kelvins) for bulk Cd\textsubscript{3}As\textsubscript{2}.}
	\label{fig:resistivityvstemperature}
\end{figure}

We now discuss how the bulk temperature is measured. Since the elevated bulk temperature will increase the bulk resistance of the superconducting bulk states as shown in Fig.~\ref{fig:resistivityvstemperature}, we measure the zero-bias resistivity across the bulk (using a lock-in amplifier) as a proxy for the temperature. This is advantageous relative to infrared-based bolometry since it does not perturb the electronic structure of the bulk, as well as due to the fact that electrical signals can be measured in ultrafast picosecond-range intervals \cite{ElectricalMeasurement}. We manufactured a Cd$_3$As$_2$ device to measure the superconducting bulk resistivity as a function of sample temperature. To this end, it is important to note the lower bounds for the dimensions of each Cd$_3$As$_2$ crystal. The goal of the device is to measure photons in the transmon frequency range, i.e. $5-7 \textrm{ GHz}$ \cite{Transmon1,Transmon2}. For a Dirac cone dispersion with Fermi velocity $v_F$, a photon of frequency $f$ is resonant with the band gap at the following band wavevector:
\begin{equation}
k = \frac{\pi f}{v_F}.
\end{equation}
Therefore, in order for resonance to exist at photon frequencies as low as 5 GHz, the maximum length of each Bloch state in reciprocal space must be $\Delta k \approx 1.6 \times 10^4 \textrm{ m}^{-1}$, thus implying that the minimum length of the Cd\textsubscript{3}As\textsubscript{2} surface along each dimension is $2\pi/\Delta k = 0.4 \textrm{ mm}$. We also assume that the depth of the lattice is limited by design constraints to a minimum value of 20 nm, since this is the minimum thickness that has been achieved with an MBE technique \cite{schumannQHE2018}. For a photon frequency of 5 GHz and crystal dimensions of 0.4 mm by 0.4 mm by 20 nm, the single-photon temperature gain is calculated by substituting the values $N = 1$, $\omega = \pi \times 10^{10} \textrm{ s}^{-1}$, and $V = 3.2 \times 10^{-15} \textrm{ m}^3$ into Eq.~\eqref{eq: final temperature} and linearizing: 
\begin{align} \label{eq: single-photon temperature gain}
\begin{split}
\Delta T &= \frac{1}{4T^3} \Big(4.13 \times 10^{-35} \textrm{ m}^3\textrm{K}^4\textrm{s}\Big) \frac{N\omega}{V} \\
&= \frac{1.0 \times 10^{-10} \textrm{ K}^4}{T^3}.
\end{split}
\end{align}
For temperatures above our minimum refrigerator temperature of 0.25 K, the temperature gain due to the absorption of a single photon is below 6.5 nK, which confirms our previous assumption that $\Delta T << T$.
\par
\begin{figure}[!tb]
	\centering
	\includegraphics[width=\linewidth]{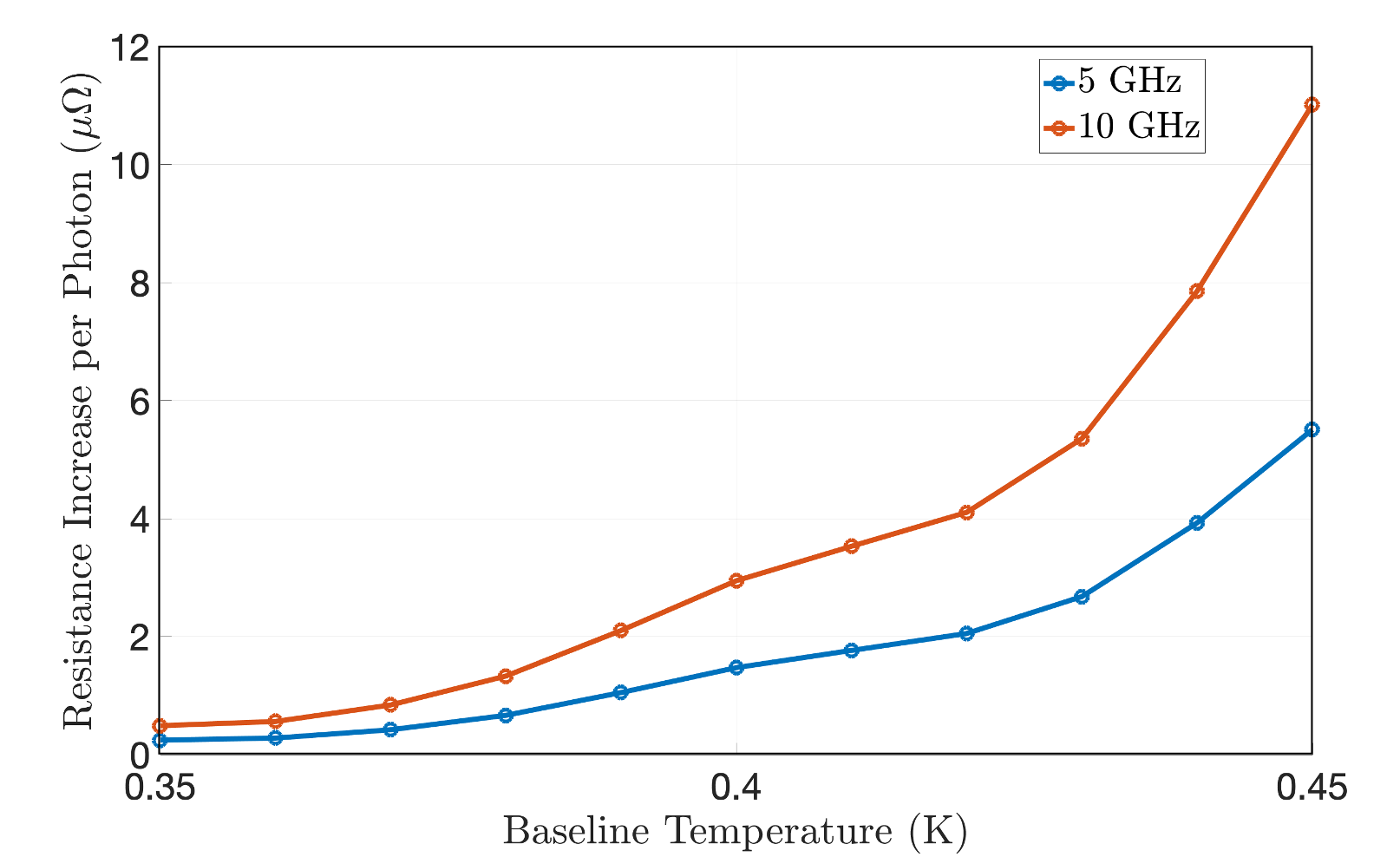}
	\caption{Plots of bulk resistance gain (in microohms) due to absorption of a single photon versus baseline temperature for photon frequencies $f = 5, 10 \textrm{ GHz}$ given sample dimensions 0.4 mm by 0.4 mm by 20 nm.}
	\label{fig:resistancegain}
\end{figure}

Finally, we use the single-photon temperature gain to determine the corresponding increase in bulk resistance. Figure~\ref{fig:resistivityvstemperature} depicts the experimental values for zero-bias resistivity as a function of temperature in bulk Cd\textsubscript{3}As\textsubscript{2} in the superconducting regime.
For temperatures above 0.35 K, the resistivity steadily increases with temperature. We will therefore use 0.35 K to 0.45 K as the range of baseline temperatures for which we will determine the single-photon bulk resistance gain. For a square lattice surface, the bulk resistance scales linearly with resistivity as $1/d$, where $d$ denotes the lattice depth. Therefore, the single-photon resistance gain relates to the slope of the resistivity with respect to temperature ($d\rho/dT$) and the single-photon temperature gain ($\Delta T$) as follows:
\begin{equation}
\Delta R = \frac{1}{d} \frac{d\rho}{dT} \Delta T
\end{equation}
For the aforementioned sample dimensions, $d = 20 \textrm{ nm}$. Substituting the expression for $\Delta T$ from Eq.~\eqref{eq: single-photon temperature gain}, we find that the single-photon resistance gain $\Delta R$ solely becomes a function of the baseline temperature $T$:
\begin{equation}
\Delta R = \frac{5.0 \times 10^{-3} \textrm{ m}^{-1}\textrm{K}^4}{T^3} \frac{d\rho}{dT}
\end{equation}
Figure~\ref{fig:resistancegain} depicts the resistance gain due to the absorption of a single photon for baseline temperatures ranging from 0.35 K to 0.45 K for the selected photon frequencies of 5 GHz and 10 GHz.
For temperatures of 0.39 K and above, the single-photon resistance gain will be greater than 1 $\mu\Omega$ for photon frequencies as low as 5 GHz, an increase which is certainly measurable using a commercially available micro-ohm meter (such as the Keysight 34420A NanoVolt/Micro-Ohm Meter by Keysight Technologies) or with a Corbino geometry sample which can even measure sub-micro-ohm resistance \cite{TsuiCorbino}. This property can therefore be exploited in order to precisely determine the number of absorbed photons for a known frequency.
\par
It is worth discussing the effect of impurities on the properties of the detector. Normally, the presence of charged impurities would lead to a shift of the Fermi level away from the Dirac point, which in turn would degrade the performance of the detector by hindering photon absorption in the microwave frequency range. However, recent experiments have demonstrated that Cd\textsubscript{3}As\textsubscript{2} is easily doped, either chemically \cite{LiuChemicalDopingCd3As2} or electrostatically \cite{LiuElectrostaticDopingCd3As2}. Therefore, the Fermi level of the surface state can be tuned so as to coincide with the Dirac point, as desired. Since the electronic structure of the surface is decoupled from that of the bulk, it is feasible to specifically dope the former while leaving the latter unaltered.
\par
Finally, we address the issue of dark count. Due to the cryogenic (sub-Kelvin) refrigerator temperature, the dark count should be negligible, as previously demonstrated for transition-edge sensors under similar temperature conditions \cite{NiwaTES}. Nonetheless, precise experimental determination of the dark count for the Cd\textsubscript{3}As\textsubscript{2} detector would serve as an important topic for future research.
\par 

\section{Discussions and Conclusion} \label{sec:conclusion}

We demonstrated a microwave photon-number resolving detector based on the topological surface states of Cd$_3$As$_2$ material. The number of photons absorbed is produced after measuring the increased resistivity of the superconducting bulk. For this, we derived in detail how much bulk temperature would elevate as a function of the absorbed number of photons and the photon frequency. We showed that the energy of the absorbed photon is rapidly transferred first to the rethermalized distribution of the surface state electron band population. Then, the electron band energy is quickly transferred to the bulk phonon modes through the deformation potential coupling. The bulk temperature is thus elevated, and finally, the superconducting bulk increases resistance, which is measured to resolve the absorbed number of photons. To address how quickly the energy is transferred from the surface electron to the bulk phonon modes, we derived the deformation potential electron-phonon coupling rate by calculating the transition matrix element and the phase space volume. As a result, we concluded that the coupling time constant ranged from nanoseconds to microseconds. Therefore, it is expected that the number of absorbed photons would be measured within several milliseconds after the absorption happens. 

Our proposed scheme accomplishes rapid photon detection based on quick (or even continuous) and accurate bulk resistance measurement. Direct measurement of the elevated temperature in bulk does not provide a feasible path due to the slow detection speed and the measurement noise in the extremely small differential temperature. It is essential to understand why the use of Cd$_3$As$_2$ bulk's semimetal feature for absorbing microwave photons is avoided. Recall that, if the baseline temperature is set above the critical temperature, the bulk's electronic bands do not open a gap, which allows the bulk electrons to be excited by the microwave photons. However, detecting the excited electron is extremely difficult for two main reasons. First, the bulk photoelectron may easily join the resistance-measuring current and be lost in the measurement process. Second, the photoelectron's energy transfer to the bulk temperature is extremely inefficient due to the reduced phase space of 3D electrons, risking the loss of photoelectrons via radiative decay rather than energy transfer to the bulk phonon modes. In contrast, the photon absorption from the surface state electrons almost surely transfers the energy to the bulk phonon modes. 

Equally important is understanding the difference between our proposed scheme and an alternative device structure of a Dirac 2D material such as graphene on the surface of a bulk superconductor. A pure graphene layer indeed does not possess a superconductor state \cite{kopnin2008bcs}, and thus can be used as a Dirac cone photon absorber of microwave photons even at a very low temperature. However, it is more difficult to fabricate this device than Cd$_3$As$_2$ which simultaneously has both bulk superconductor and surface states. In addition, the hybrid structure suffers from inefficient electronic energy transfer to the bulk phonons due to the mismatch of lattice constants. Instead, as previous research on graphene single-photon detectors has shown, the inefficient electronic energy transfer to phonons is used for efficient capture of the photoelectron in the electrodes \cite{walsh2017graphene}. However, in this case, the photon-number resolving feature is lost. In comparison, our scheme utilizes the surface state electrons of Cd$_3$As$_2$ as a microwave photon absorber and the bulk superconductor of the same material for detecting the number of photons absorbed. The distinct advantage of our method is to provide a deterministic photon-number resolving capability in microwave photon detection.

It is also worth understanding the advantage that our scheme offers over traditional transition-edge-sensor (TES) based detectors. Due to the need for a significant voltage bias in measuring the resistance of the TES bulk \cite{GerritsTES}, a large source-drain current is generated, causing undesired side effects such as flicker noise. A Cd\textsubscript{3}As\textsubscript{2}-based detector avoids this issue by enabling zero-bias resistance measurement.

We now discuss the design strategy of maximizing the photon absorption probability of the device. Note that each crystal surface features an absorption rate of 0.3-0.6\%. Therefore, it is possible to have a near unity quantum efficiency if about 2000 bulk crystal layers are vertically stacked in a heterostructure (such that they are in series from the point of view of the incoming photon), while measuring the bulk zero-bias resistivity for each of the crystals separately. With the advent of more advanced manufacturing techniques, such heterostructure is increasingly becoming possible \cite{tongay2014tuning}. Another means of achieving the same goal is by placing a single-layer detector in an optical cavity bounded by high-reflectivity mirrors. Since Bragg mirrors can feature transmittance rates as low as 1 ppm \cite{BraggMirrorsMinimumTransmittance}, the total probability that a photon is lost through one of the mirrors will be negligible even after thousands of round trips through the cavity, thus ensuring a near-unity detector efficiency.

\begin{acknowledgements}

Sandia National Laboratories is a multimission laboratory managed and operated by National Technology \& Engineering Solutions of Sandia, LLC, a wholly owned subsidiary of Honeywell International Inc., for the U.S. Department of Energy’s National Nuclear Security Ad- ministration under Contract No. DE-NA-0003525.

\end{acknowledgements}

\appendix

\section{Matrix Elements for Interaction Between Surface Electrons and Bulk Phonons} \label{sec: Matrix Elements for Interaction Between Surface Electrons and Bulk Phonons}
In this section, we estimate the matrix elements corresponding to interaction between the surface electron modes and bulk phonon modes by building from the analogous elements for the bulk electron-phonon interaction. Labeling the direction perpendicular to the surface as $\hat{z}$, we  
represent the surface state wavefunction as a product of the $xy$-plane wavefunction and a pulse-like function of $z$:
\begin{equation}
\Psi(x,y,z) = \phi(x,y) \psi(z),
\end{equation}
where $\psi(z)$ is expressed such that its amplitude-squared becomes a broadened Dirac-delta function with a width of $a$:
\begin{equation}
\psi(z) =
\begin{cases}
\frac{1}{\sqrt{a}}, & 0 < z < a \\
0, & \text{otherwise}
\end{cases}.
\end{equation}
Here, $a$ denotes the approximate width of the surface state.
\par
Next, we seek to express a surface mode as a superposition of bulk modes by decomposing $\psi(z)$ into a superposition of modes with well-defined $z$-wavevector $k_z$:
\begin{equation}
\ket{\psi} = c(k_z) \ket{k_z},
\end{equation}
where $\ket{k_z}$ denotes the plane wave state with wavevector $k_z$, taking the following form with respect to the lattice depth $d$ when projected onto the position space:
\begin{equation}
\braket{z|k_z} = \frac{1}{\sqrt{d}} e^{i k_z z}.
\end{equation}
From the Heisenberg uncertainty principle, we intuitively know that the range of $z$-direction momentum is approximated as $\Delta p_z \approx h/a$, leading to a wavevector range of $\Delta k_z \approx 2\pi/a$. We quantitatively determine the superposition coefficients $c(k_z)$ as follows:
\begin{align}
\begin{split}
c(k_z) &= \braket{k_z|\psi} \\
&= \int dz \braket{k_z|z} \braket{z|\psi} \\
&= \frac{1}{\sqrt{ad}} \int_{0}^{a} dz e^{-i k_z z} \\
&= \frac{i}{\sqrt{ad}} \frac{e^{-i k_z a} - 1}{k_z}.
\end{split}
\end{align}
For low values of $k_z a$ (i.e., $k_z a \lessapprox 1$), the coefficient can be estimated as $\sqrt{a/d}$. Since the span of each plane-wave state in reciprocal space is $2\pi/d$, this accords with the intuition that the overall reciprocal space in the $z$-direction spans a length of $2\pi/a$, subdivided into $a/d$ wavevectors with a roughly uniform superposition coefficient for each. 
\par 
The intraband matrix element corresponding to the emission of a phonon of wavevector $\bm{q}$ in branch $\mu$ by a surface electron can thus be expressed in terms of the analogous matrix elements for bulk electrons:
\begin{widetext}
\begin{align} \label{eq: H_emit}
\begin{split}
H_{emit}(\bm{k_{xy}},\bm{q}) &= \braket{\bm{k_{xy} - q_{xy}}, \psi, n_{\mu,\bm{q}} + 1|\hbar g_{\mu,\bm{k},\bm{q}} c^{\dag}_{\bm{k-q}} c_{\bm{k}} b^{\dag}_{\mu,\bm{q}}|\bm{k_{xy}}, \psi, n_{\mu,\bm{q}}} \\
&= \sum_{k_z} \braket{\psi|k_z - q_z} \braket{\bm{k - q}, n_{\mu,\bm{q}} + 1|\hbar g_{\mu,\bm{k},\bm{q}} c^{\dag}_{\bm{k-q}} c_{\bm{k}} b^{\dag}_{\mu,\bm{q}}|\bm{k}, n_{\mu,\bm{q}}} \braket{k_z|\psi} \\
&\approx \frac{a}{d} \sum_{k_z = -\frac{\pi}{a}}^{\frac{\pi}{a}} \braket{\bm{k - q}, n_{\mu,\bm{q}} + 1|\hbar g_{\mu,\bm{k},\bm{q}} c^{\dag}_{\bm{k-q}} c_{\bm{k}} b^{\dag}_{\mu,\bm{q}}|\bm{k}, n_{\mu,\bm{q}}}.
\end{split}
\end{align}
\end{widetext}
Note that this approximation is valid specifically if the maximum amplitude of the emitted phonon wavevector is much smaller than the maximum amplitude of $k_z$, i.e. if $q_{max} << \pi/a$, which holds true for long-wavelength acoustic phonons in the linear dispersion regime.
\par 
As a final step in the generic matrix element calculation, we can show that the carrier-phonon matrix element is exactly invariant in the initial carrier wavevector $k_z$, since each $k_z$ corresponds to a plane-wave state with well-defined $z$-momentum $\hbar k_z$. Specifically, the matrix element of a spatial function $f(\bm{r})$ connecting an initial carrier state $\ket{\bm{k}}$ to a final state $\ket{\bm{k-q}}$ (where the wavevectors are three-dimensional) is simplified as follows:
\begin{widetext}
\begin{align}
\begin{split}
\braket{\bm{k-q}|f(\bm{r})|\bm{k}} &= \braket{\bm{k_{xy}-q_{xy}},k_z-q_z|f(\bm{r})|\bm{k_{xy}},k_z} \\
&= \int d^3r \phi^*_{\bm{k_{xy}-q_{xy}}}(x,y) \frac{1}{\sqrt{d}} e^{-i (k_z - q_z) z} f(\bm{r}) \phi_{\bm{k_{xy}}}(x,y) \frac{1}{\sqrt{d}} e^{i k_z z} \\
&= \frac{1}{d} \int d^3r e^{i q_z z} \phi^*_{\bm{k_{xy}-q_{xy}}}(x,y) f(\bm{r}) \phi_{\bm{k_{xy}}}(x,y).
\end{split}
\end{align}
\end{widetext}
As a result, the terms in the summation in Eq.~\eqref{eq: H_emit} are equivalent. Since the total number of valid plane-wave wavevectors in the summation is $d/a$ (as previously discussed), the matrix element $H_{emit}$ becomes invariant in the lattice depth $d$, as desired:
\begin{align}
\begin{split}
&H_{emit}(\bm{k_{xy}},\bm{q}) \\
&\approx \frac{a}{d} \frac{d}{a} \braket{\bm{k - q}, n_{\mu,\bm{q}} + 1|\hbar g_{\mu,\bm{k},\bm{q}} c^{\dag}_{\bm{k-q}} c_{\bm{k}} b^{\dag}_{\mu,\bm{q}}|\bm{k}, n_{\mu,\bm{q}}} \\
&\approx \braket{\bm{k - q}, n_{\mu,\bm{q}} + 1|\hbar g_{\mu,\bm{k},\bm{q}} c^{\dag}_{\bm{k-q}} c_{\bm{k}} b^{\dag}_{\mu,\bm{q}}|\bm{k}, n_{\mu,\bm{q}}}.
\end{split}
\end{align}
Therefore, the electron-phonon matrix element $M_{\bm{k_{xy}},\bm{p_{xy}}}^{\mu,\bm{q}}$ corresponding to the transition from an initial electronic wavevector $\bm{k_{xy}}$ to a final wavevector $\bm{p_{xy}}$ via the emission of a phonon of wavevector $\bm{q}$ in branch $\mu$ is equivalent to the bulk matrix element $M_{\bm{k},\bm{p}}^{\mu,\bm{q}}$, where $k_z = 0$ and $p_z = -q_z$.

\section{Calculating the Surface Electron Cooling Rate} \label{sec: Calculating the Surface Electron Cooling Rate}
We start by representing the cooling rate of the surface electrons as the following summation over initial electron wavevectors $\bm{k}$, final electron wavevectors $\bm{p}$, and phonon wavevectors $\bm{q}$, with the direction of $\bm{k}$ defined as the $x$-axis and the projection of $\bm{p}$ on the $xy$-plane labeled as $p_{xy}$:
\begin{align}
\begin{split}
Q &= \frac{2\pi}{\hbar} \sum_{\bm{k},\bm{p},\bm{q}} \hbar v_F \Big(k - p_{xy}\Big) \frac{C}{V} q \Big(f(\hbar v_F k) - f(\hbar v_F p_{xy})\Big) \\
&\quad \times \Big(n_{T_L}(v_s q) - n_{T_e}(v_s q)\Big) \\
&\quad \times \delta_{\bm{k},\bm{p+q}} \delta\Big(\hbar v_F k - \hbar v_F p_{xy} - \hbar v_s q\Big).
\end{split}
\end{align}
Given an overall lattice volume $V$ and surface state area $A$, the discrete summation over $\bm{k}$ and $\bm{p}$ can be converted to integrals as follows:
\begin{align}
\begin{split}
\sum_{\bm{k}} &\rightarrow \frac{A}{(2\pi)^2} \int d^2k, \\
\sum_{\bm{p}} &\rightarrow \frac{V}{(2\pi)^3} \int d^3p = \frac{V}{(2\pi)^3} \int dp_{xy} p_{xy} \int d\theta_{\bm{p}} \int dp_z.
\end{split}
\end{align}
Also, the Dirac delta can be re-written as a function of $p_{xy}$:
\begin{equation}
\delta\Big(\hbar v_F k - \hbar v_F p_{xy} - \hbar v_s q\Big) = \frac{1}{\hbar v_F} \delta\bigg(p_{xy} - k + \frac{v_s}{v_F}q\bigg).
\end{equation}
Therefore, the Dirac delta collapses the integral over $p_{xy}$ to $p_{xy} = k - v_s q/v_F$. The rate $Q$ reduces to the following form:
\begin{widetext}
\begin{align}
\begin{split}
Q &= \frac{2\pi}{\hbar} \frac{A}{(2\pi)^2} \frac{V}{(2\pi)^3} \\ &\quad \times \sum_{\bm{q}} \int d^2k \int d\theta_{\bm{p}} \int dp_z \bigg(k - \frac{v_s}{v_F}q\bigg) \frac{v_s}{v_F} q \frac{C}{V} q \Big(f(\hbar v_F k) - f(\hbar v_F k - \hbar v_s q)\Big) \Big(n_{T_L}(v_s q) - n_{T_e}(v_s q)\Big) \delta_{\bm{k},\bm{p+q}}.
\end{split}
\end{align}
\end{widetext}
Next, we apply the Kronecker delta, which restricts the summation over phonon wavevectors to $\bm{q} = \bm{k-p}$. Also, as discussed in the main text, we can apply the approximation $p \approx p_z >> k$ for nearly all of the phase space. This argument is further strengthened by the fact that the integrand approaches zero for low values of $q$ for which this approximation is the weakest. Therefore, we set $q = p_z$. Since all terms in the integrand are now functions solely of $k$ or $p_z$, we have collapsed the integral over $\bm{p}$ to an integral over $p_z$ multiplied by the circumference of the circle generated by making a plane cut (along the $p_xp_y$ plane) through the cone. Also, the fact that the radius of the cone at a generic value of $p_z$ equals $k - (v_s/v_F)p_z$ implies that the circumference of the aforementioned circle is $2\pi (k - (v_s/v_F)q_z)$ and the upper bound of $q_z$ (corresponding to the cone vertex) is $(v_F/v_s)k$. Therefore, $Q$ simplifies to the following:
\begin{widetext}
\begin{align}
\begin{split}
Q &\approx 2 \frac{2\pi}{\hbar} \frac{A}{(2\pi)^2} \frac{V}{(2\pi)^3} \frac{C}{V} \frac{v_s}{v_F} \\
&\quad \times \int_0^{\infty} dk 2\pi k \int_0^{\frac{v_F}{v_s}k} dp_z 2\pi \bigg(k - \frac{v_s}{v_F}p_z\bigg) p_z^2 \Big(f(\hbar v_F k) - f(\hbar v_F k - \hbar v_s p_z)\Big) \Big(n_{T_L}(v_s p_z) - n_{T_e}(v_s p_z)\Big).
\end{split}
\end{align}
\end{widetext}
Note that the factor of 2 in front derives from the fact that the available phase space is actually a double cone, extending in both the $+$ and $-$ directions along the $p_z$-axis.
\par 
We now examine the parenthetical term corresponding to the net phonon number. In the limit $\Delta T << T_e, T_L$, where $\Delta T = T_e - T_L$, we solve for a generic value of $T \approx T_e \approx T_L$ to first order in $\Delta T$:
\begin{align}
\begin{split}
&n_{T_L}(v_s p_z) - n_{T_e}(v_s p_z) \\
&= \bigg(e^{\frac{\hbar v_s p_z}{k_B T_L}} - 1\bigg)^{-1} - \bigg(e^{\frac{\hbar v_s p_z}{k_B T_e}} - 1\bigg)^{-1} \\
&\approx \frac{e^{\frac{\hbar v_s p_z}{k_B T}}}{\Big(e^{\frac{\hbar v_s p_z}{k_B T_L}} - 1\Big)^2} \frac{\hbar v_s p_z \Delta T}{k_B T^2}.
\end{split}
\end{align}
The exponential term will set an approximate upper bound for the emitted phonon energy. For low temperatures (e.g. $T = 0.45 \textrm{ K}$), this ensures that the emitted phonons fall within the long-wavelength acoustic mode limit. 
\par 
Next, we write out the expression for the electron occupation number difference between the initial and final states:
\begin{align}
\begin{split}
&f(\hbar v_F k) - f(\hbar v_F k - \hbar v_s q_z) \\
&= \frac{1}{e^{\frac{\hbar v_F k}{k_B T}} + 1} - \frac{1}{e^{\frac{\hbar v_F k - \hbar v_s p_z}{k_B T}} + 1}.
\end{split}
\end{align}
We are now in a position to simplify the double integral by defining the variables $x = \hbar v_F k/(k_B T)$ and $y = \hbar v_s p_z/(k_B T)$, yielding the following expression for $Q$:
\begin{widetext}
\begin{align} \label{eq: intraband Q}
\begin{split}
Q &\approx \frac{4\pi C}{\hbar} \frac{A}{(2\pi)^5} \frac{v_s}{v_F} \frac{\hbar v_s \Delta T}{k_B T^2} 4\pi^2 \\
&\quad \times \int_0^{\infty} dk k \int_0^{\frac{v_F}{v_s}k} dp_z \bigg(k - \frac{v_s}{v_F}p_z\bigg) p_z^3 \frac{e^{\frac{\hbar v_s p_z}{k_B T}}}{\Big(e^{\frac{\hbar v_s p_z}{k_B T_L}} - 1\Big)^2} \Bigg(\frac{1}{e^{\frac{\hbar v_F k}{k_B T}} + 1} - \frac{1}{e^{\frac{\hbar v_F k - \hbar v_s p_z}{k_B T}} + 1}\Bigg) \\
&= \frac{AC}{2\pi^2 \hbar} \frac{v_s}{v_F} \bigg(\frac{\hbar v_s}{k_B T}\bigg) \frac{\Delta T}{T} \bigg(\frac{k_B T}{\hbar v_F}\bigg)^3 \bigg(\frac{k_B T}{\hbar v_s}\bigg)^4 \\
&\quad \times \int_0^{\infty} dx x \int_0^x dy (x-y) y^3 \frac{e^y}{(e^y - 1)^2} \Bigg(\frac{1}{e^{x} + 1} - \frac{1}{e^{x-y} + 1}\Bigg).
\end{split}
\end{align}
\end{widetext}
Note that our analysis so far has been restricted to the case of intraband carrier transitions through carrier-phonon interaction. Unlike the case of bulk carriers, where the band structure prohibits interband carrier-phonon scattering, the carrier dispersion for the surface state allows such scattering. The phase space area for this interaction can be constructed by merging the conical phase space for the electron-phonon interaction with the corresponding cone for the hole-phonon interaction, with the pair of cones meeting at $p_z = v_F k/v_s$. Just like we restricted the range of values for $p_z$ for the intraband scattering to $p_z < v_F k/v_s$, we will restrict the range for interband scattering to $p_z > v_F k/v_s$. Recall that the amplitude-squared of the carrier-phonon matrix element for Cd$_3$As$_2$ is identical for interband and intraband interaction except for the following proportionality \cite{LundgrenFiete}:
\begin{equation}
\sum_{\mu} \Big|M_{\bm{k},\bm{p}}^{\mu,\bm{q}}\Big|^2 \propto 1 + s \cos{\theta},
\end{equation}
where $\theta$ represents the angle between $\bm{k}$ and $\bm{p}$, and $s = 1, -1$ for intraband and interband interactions, respectively. As mentioned previously, $p_z >> k$ for nearly all of the phase space area for the intraband interactions, and since the interband interactions involve even higher values for $p_z$, this is clearly true for such processes as well. Therefore, for both intraband and interband interactions, we can make the approximation $\cos{\theta} \approx 0$, causing the interband and intraband matrix element amplitude-squared values to scale linearly with the phonon wavevector in the same manner. In order to construct the equivalent of Eq.~\eqref{eq: intraband Q} for interband carrier-phonon scattering, the only changes we make are thus the range of $p_z$, as previously mentioned, as well as the term in the integrand corresponding to the cone radius, which flips as $k - v_s p_z/v_F \rightarrow v_s p_z/v_F - k$:
\begin{widetext}
	\begin{align} \label{eq: interband Q}
	\begin{split}
	Q_{inter} &\approx \frac{4\pi C}{\hbar} \frac{A}{(2\pi)^5} \frac{v_s}{v_F} \frac{\hbar v_s \Delta T}{k_B T^2} 4\pi^2 \\
	&\quad \times \int_0^{\infty} dk k \int_{\frac{v_F}{v_s}k}^{\infty} dp_z \bigg(\frac{v_s}{v_F}p_z - k\bigg) p_z^3 \frac{e^{\frac{\hbar v_s p_z}{k_B T}}}{\Big(e^{\frac{\hbar v_s p_z}{k_B T_L}} - 1\Big)^2} \Bigg(\frac{1}{e^{\frac{\hbar v_F k}{k_B T}} + 1} - \frac{1}{e^{\frac{\hbar v_F k - \hbar v_s p_z}{k_B T}} + 1}\Bigg) \\
	&= \frac{AC}{2\pi^2 \hbar} \frac{v_s}{v_F} \bigg(\frac{\hbar v_s}{k_B T}\bigg) \frac{\Delta T}{T} \bigg(\frac{k_B T}{\hbar v_F}\bigg)^3 \bigg(\frac{k_B T}{\hbar v_s}\bigg)^4 \\
	&\quad \times \int_0^{\infty} dx x \int_x^{\infty} dy (y-x) y^3 \frac{e^y}{(e^y - 1)^2} \Bigg(\frac{1}{e^{x} + 1} - \frac{1}{e^{x-y} + 1}\Bigg).
	\end{split}
	\end{align}
\end{widetext}

\end{document}